\documentclass[aps,prl,twocolumn,superscriptaddress,showpacs]{revtex4} 

\usepackage{graphicx}
\usepackage{rotating}

\begin{document}


\newcommand{\fix}[1]{} 
\newcommand{\MET}{\mbox{$\raisebox{.3ex}{$\not\!$}E_T$}}
\newcommand{\METVEC}{\mbox{$\raisebox{.3ex}{$\not\!$}{\vec E}_T$}}
\newcommand{\MPTVEC}{\mbox{$\raisebox{.3ex}{$\not\!$}{\vec {p}_T}$}}

\def\DZero{D0~}

\newcommand{\Et}{\mbox{$E_T$}}
\newcommand{\Pt}{\mbox{$p_T$}}
\newcommand{\Ht}{\mbox{$H_T$}}
\newcommand{\met}{\mbox{$\protect \raisebox{0.3ex}{$\not$}\Et$}}
\newcommand{\mpt}{\mbox{$\protect \raisebox{0.3ex}{$\not$}\Pt$}}



\title{First Observation of Electroweak Single Top Quark Production}
   

\affiliation{Institute of Physics, Academia Sinica, Taipei, Taiwan 11529, Republic of China} 
\affiliation{Argonne National Laboratory, Argonne, Illinois 60439} 
\affiliation{University of Athens, 157 71 Athens, Greece} 
\affiliation{Institut de Fisica d'Altes Energies, Universitat Autonoma de Barcelona, E-08193, Bellaterra (Barcelona), Spain} 
\affiliation{Baylor University, Waco, Texas  76798} 
\affiliation{Istituto Nazionale di Fisica Nucleare Bologna, $^x$University of Bologna, I-40127 Bologna, Italy} 
\affiliation{Brandeis University, Waltham, Massachusetts 02254} 
\affiliation{University of California, Davis, Davis, California  95616} 
\affiliation{University of California, Los Angeles, Los Angeles, California  90024} 
\affiliation{University of California, San Diego, La Jolla, California  92093} 
\affiliation{University of California, Santa Barbara, Santa Barbara, California 93106} 
\affiliation{Instituto de Fisica de Cantabria, CSIC-University of Cantabria, 39005 Santander, Spain} 
\affiliation{Carnegie Mellon University, Pittsburgh, PA  15213} 
\affiliation{Enrico Fermi Institute, University of Chicago, Chicago, Illinois 60637}
\affiliation{Comenius University, 842 48 Bratislava, Slovakia; Institute of Experimental Physics, 040 01 Kosice, Slovakia} 
\affiliation{Joint Institute for Nuclear Research, RU-141980 Dubna, Russia} 
\affiliation{Duke University, Durham, North Carolina  27708} 
\affiliation{Fermi National Accelerator Laboratory, Batavia, Illinois 60510} 
\affiliation{University of Florida, Gainesville, Florida  32611} 
\affiliation{Laboratori Nazionali di Frascati, Istituto Nazionale di Fisica Nucleare, I-00044 Frascati, Italy} 
\affiliation{University of Geneva, CH-1211 Geneva 4, Switzerland} 
\affiliation{Glasgow University, Glasgow G12 8QQ, United Kingdom} 
\affiliation{Harvard University, Cambridge, Massachusetts 02138} 
\affiliation{Division of High Energy Physics, Department of Physics, University of Helsinki and Helsinki Institute of Physics, FIN-00014, Helsinki, Finland} 
\affiliation{University of Illinois, Urbana, Illinois 61801} 
\affiliation{The Johns Hopkins University, Baltimore, Maryland 21218} 
\affiliation{Institut f\"{u}r Experimentelle Kernphysik, Universit\"{a}t Karlsruhe, 76128 Karlsruhe, Germany} 
\affiliation{Center for High Energy Physics: Kyungpook National University, Daegu 702-701, Korea; Seoul National University, Seoul 151-742, Korea; Sungkyunkwan University, Suwon 440-746, Korea; Korea Institute of Science and Technology Information, Daejeon, 305-806, Korea; Chonnam National University, Gwangju, 500-757, Korea} 
\affiliation{Ernest Orlando Lawrence Berkeley National Laboratory, Berkeley, California 94720} 
\affiliation{University of Liverpool, Liverpool L69 7ZE, United Kingdom} 
\affiliation{University College London, London WC1E 6BT, United Kingdom} 
\affiliation{Centro de Investigaciones Energeticas Medioambientales y Tecnologicas, E-28040 Madrid, Spain} 
\affiliation{Massachusetts Institute of Technology, Cambridge, Massachusetts  02139} 
\affiliation{Institute of Particle Physics: McGill University, Montr\'{e}al, Qu\'{e}bec, Canada H3A~2T8; Simon Fraser University, Burnaby, British Columbia, Canada V5A~1S6; University of Toronto, Toronto, Ontario, Canada M5S~1A7; and TRIUMF, Vancouver, British Columbia, Canada V6T~2A3} 
\affiliation{University of Michigan, Ann Arbor, Michigan 48109} 
\affiliation{Michigan State University, East Lansing, Michigan  48824}
\affiliation{Institution for Theoretical and Experimental Physics, ITEP, Moscow 117259, Russia} 
\affiliation{University of New Mexico, Albuquerque, New Mexico 87131} 
\affiliation{Northwestern University, Evanston, Illinois  60208} 
\affiliation{The Ohio State University, Columbus, Ohio  43210} 
\affiliation{Okayama University, Okayama 700-8530, Japan} 
\affiliation{Osaka City University, Osaka 588, Japan} 
\affiliation{University of Oxford, Oxford OX1 3RH, United Kingdom} 
\affiliation{Istituto Nazionale di Fisica Nucleare, Sezione di Padova-Trento, $^y$University of Padova, I-35131 Padova, Italy} 
\affiliation{LPNHE, Universite Pierre et Marie Curie/IN2P3-CNRS, UMR7585, Paris, F-75252 France} 
\affiliation{University of Pennsylvania, Philadelphia, Pennsylvania 19104}
\affiliation{Istituto Nazionale di Fisica Nucleare Pisa, $^z$University of Pisa, $^{aa}$University of Siena and $^{bb}$Scuola Normale Superiore, I-56127 Pisa, Italy} 
\affiliation{University of Pittsburgh, Pittsburgh, Pennsylvania 15260} 
\affiliation{Purdue University, West Lafayette, Indiana 47907} 
\affiliation{University of Rochester, Rochester, New York 14627} 
\affiliation{The Rockefeller University, New York, New York 10021} 
\affiliation{Istituto Nazionale di Fisica Nucleare, Sezione di Roma 1, $^{cc}$Sapienza Universit\`{a} di Roma, I-00185 Roma, Italy} 

\affiliation{Rutgers University, Piscataway, New Jersey 08855} 
\affiliation{Texas A\&M University, College Station, Texas 77843} 
\affiliation{Istituto Nazionale di Fisica Nucleare Trieste/Udine, I-34100 Trieste, $^{dd}$University of Trieste/Udine, I-33100 Udine, Italy} 
\affiliation{University of Tsukuba, Tsukuba, Ibaraki 305, Japan} 
\affiliation{Tufts University, Medford, Massachusetts 02155} 
\affiliation{Waseda University, Tokyo 169, Japan} 
\affiliation{Wayne State University, Detroit, Michigan  48201} 
\affiliation{University of Wisconsin, Madison, Wisconsin 53706} 
\affiliation{Yale University, New Haven, Connecticut 06520} 
\author{T.~Aaltonen}
\affiliation{Division of High Energy Physics, Department of Physics, University of Helsinki and Helsinki Institute of Physics, FIN-00014, Helsinki, Finland}
\author{J.~Adelman}
\affiliation{Enrico Fermi Institute, University of Chicago, Chicago, Illinois 60637}
\author{T.~Akimoto}
\affiliation{University of Tsukuba, Tsukuba, Ibaraki 305, Japan}
\author{B.~\'{A}lvarez~Gonz\'{a}lez$^s$}
\affiliation{Instituto de Fisica de Cantabria, CSIC-University of Cantabria, 39005 Santander, Spain}
\author{S.~Amerio$^y$}
\affiliation{Istituto Nazionale di Fisica Nucleare, Sezione di Padova-Trento, $^y$University of Padova, I-35131 Padova, Italy} 

\author{D.~Amidei}
\affiliation{University of Michigan, Ann Arbor, Michigan 48109}
\author{A.~Anastassov}
\affiliation{Northwestern University, Evanston, Illinois  60208}
\author{A.~Annovi}
\affiliation{Laboratori Nazionali di Frascati, Istituto Nazionale di Fisica Nucleare, I-00044 Frascati, Italy}
\author{J.~Antos}
\affiliation{Comenius University, 842 48 Bratislava, Slovakia; Institute of Experimental Physics, 040 01 Kosice, Slovakia}
\author{G.~Apollinari}
\affiliation{Fermi National Accelerator Laboratory, Batavia, Illinois 60510}
% ...

\collaboration{CDF Collaboration\footnote{With visitors from $^a$University of Massachusetts Amherst, Amherst, Massachusetts 01003,
$^b$Universiteit Antwerpen, B-2610 Antwerp, Belgium, 
$^c$University of Bristol, Bristol BS8 1TL, United Kingdom,
$^d$Chinese Academy of Sciences, Beijing 100864, China, 
$^e$Istituto Nazionale di Fisica Nucleare, Sezione di Cagliari, 09042 Monserrato (Cagliari), Italy,
$^f$University of California Irvine, Irvine, CA  92697, 
$^g$University of California Santa Cruz, Santa Cruz, CA  95064, 
$^h$Cornell University, Ithaca, NY  14853, 
$^i$University of Cyprus, Nicosia CY-1678, Cyprus, 
$^j$University College Dublin, Dublin 4, Ireland,
$^k$University of Edinburgh, Edinburgh EH9 3JZ, United Kingdom, 
$^l$University of Fukui, Fukui City, Fukui Prefecture, Japan 910-0017
$^m$Kinki University, Higashi-Osaka City, Japan 577-8502
$^n$Universidad Iberoamericana, Mexico D.F., Mexico,
$^o$Queen Mary, University of London, London, E1 4NS, England,
$^p$University of Manchester, Manchester M13 9PL, England, 
$^q$Nagasaki Institute of Applied Science, Nagasaki, Japan, 
$^r$University of Notre Dame, Notre Dame, IN 46556,
$^s$University de Oviedo, E-33007 Oviedo, Spain, 
$^t$Texas Tech University, Lubbock, TX  79609, 
$^u$IFIC(CSIC-Universitat de Valencia), 46071 Valencia, Spain,
$^v$University of Virginia, Charlottesville, VA  22904,
$^w$Bergische Universit\"at Wuppertal, 42097 Wuppertal, Germany,
$^{ee}$On leave from J.~Stefan Institute, Ljubljana, Slovenia, 
}}
\noaffiliation

\date{March 4, 2009}

\begin{abstract}
We report the first observation of single top quark production using
3.2~fb$^{-1}$ of $p{\bar{p}}$ collision data with $\sqrt{s}=1.96$~TeV
collected by the Collider Detector at Fermilab.  The significance of
the observed data is 5.0 standard deviations, and the expected
sensitivity for standard model production and decay is in excess of
5.9 standard deviations.  Assuming $m_t=175$~GeV/$c^2$, we measure a
cross section of $2.3^{+0.6}_{-0.5}(\mathrm{stat+syst})$~pb, extract
the CKM matrix element value \mbox{$|V_{tb}|=0.91 \pm 0.11 (\mathrm{stat+syst})\pm 0.07(\mathrm{theory})$}, and set the limit
$|V_{tb}|>0.71$ at the 95\% C.L.

\end{abstract}




\maketitle


In the standard model (SM), top quarks are expected to be produced
singly in $p{\bar p}$ collisions through $s$-channel or $t$-channel
exchange of a virtual $W$ boson~\cite{willen}.  The reasons for
studying single top quarks are compelling: the production cross
section is directly proportional to the square of the CKM
matrix~\cite{ckm} element $|V_{tb}|$, and thus a measurement of the
rate constrains fourth-generation models, models with flavor-changing
neutral currents, and other new phenomena~\cite{Tait:2000sh}.
Electroweak production of single top quarks is a difficult process to
measure because the expected expected production cross section for the
combined s- and t-channels
($\sigma_{st}\sim2.9$~pb~\cite{harris,stxs}) is much smaller than
those of competing background processes, and it is also smaller than
the uncertainty on the total background rate.  The presence of only
one top quark in the event provides fewer features to use in
separating the signal from background, compared with measurements of
top pair production ($t{\bar{t}}$), which was first observed in
1995~\cite{ttbar}.

To overcome these challenges, a variety of multivariate techniques for
separating single top events from the backgrounds have been developed.
Using different combinations of techniques, both the CDF and D0
collaborations have published evidence for single top quark production
at significance levels of 3.7 and 3.6 standard deviations,
respectively~\cite{CDF,d0evidence}.  The analysis described in this
Letter supersedes that of Ref.~\cite{CDF} and achieves a significantly
improved sensitivity by including a larger data sample and by adding
three new analyses.  We report a signal significance of 5.0 standard
deviations, thus conclusively observing electroweak production of
single top quarks, and we make the most precise measurement of
$|V_{tb}|$ to date.


We assume that single top quarks are produced in the $s$- and
$t$-channel modes with the SM ratio, and that the branching ratio of
the top quark to $Wb$ is 100\%.  We seek events in which the $W$ boson
decays leptonically in order to improve the signal-to-background ratio
$s/b$.  We simulate single top events using the tree-level
matrix-element generator {\sc mad\-event}~\cite{madevent}.  The
$t$-channel signal is modeled by the two processes $qb\rightarrow
q^\prime t$ and $qg\rightarrow q^\prime t{\bar b}$, which are
combined to match the event kinematics predicted by a fully
differential NLO calculation~\cite{stxs,janlueck}.

  A total of six analyses are combined to yield the final results
reported here.  The likelihood function (LF), matrix element (ME), and
neural network (NN) analyses of~\cite{CDF} are re-used with an
additional 1~fb$^{-1}$ of integrated luminosity; their methods remain
unchanged.  The three new analyses introduced here are: a boosted
decision tree (BDT), a likelihood function optimized for $s$-channel
single top production (LFS), and a neural-network-based analysis of
events with missing transverse energy \MET~\cite{coord} and jets (MJ).
The BDT and LFS analyses use events that overlap with the LF, ME, and
NN analyses, while the MJ analysis uses an orthogonal event selection
that adds about 30\% to the signal acceptance.  This paper
concentrates on the three new analyses and their combination with the
analyses of~\cite{CDF} using 3.2 fb$^{-1}$ of integrated luminosity
collected with the CDF II detector~\cite{CDFdet}.



For the LF, ME, NN, BDT, and LFS analyses we select
$\ell~+~$\MET~+~jet events as described in~\cite{CDF}, where
$\ell$ is an explicitly reconstructed electron or muon from the $W$ boson
decay and at least one jet is identified as containing a $B$ hadron.
The background has contributions from events in which a $W$ boson is
produced in association with one or more heavy-flavor jets ($W+HF$),
events with mistakenly $b$-tagged light-flavor jets (mistags), multijet
events (QCD), $t{\bar{t}}$ and diboson processes, as well as $Z$+jet
events.  The expected event yields in Table~\ref{tab:events} are
estimated as in~\cite{CDF} where the signal, $t{\bar{t}}$, and
diboson categories are Monte Carlo (MC) predictions scaled to the total
integrated luminosity while the remaining categories use 
predictions derived from data control samples.
The uncertainties quoted in Table~\ref{tab:events} include theoretical
uncertainties, the luminosity uncertainty for the MC predictions,
and experimental uncertainties for the data-driven background
normalizations.

  The MJ analysis is designed to select events with \protect\MET\ and
jets and to veto events selected by the $\ell+\MET$+jet analyses.  It
accepts events in which the $W$ boson decays into $\tau$ leptons and
those in which the electron or muon fails the lepton identification
criteria.  We use data corresponding to 2.1 fb$^{-1}$ of integrated
luminosity for the MJ analysis and select events that have $\MET >
50$~GeV and two jets within $|\eta|<2.0$, at least one of which has
$|\eta|<0.9$.  The jet energy measurements include information from
both the calorimeter and the charged-particle spectrometer.  Events
must have one jet with transverse energy $E_T$ greater than 35~GeV,
and a second jet with $E_T$ greater than 25~GeV.  The angular
separation between the two jets, $\Delta
R=\sqrt{(\Delta\eta)^2+(\Delta\phi)^2}$, is required to exceed 1.0.
We reject events with four or more jets with $E_T>15$~GeV in
$|\eta|<2.4$ in order to reduce the multijet (QCD) and $t{\bar{t}}$
backgrounds.  We identify $b$ jets with the same algorithm used 
in~\cite{CDF} supplemented with a jet probability
algorithm~\cite{jetprob}.

  The primary background in the MJ analysis is QCD events
  in which mismeasured jet energies produce large~\METVEC\
  aligned in the same direction as jets. To reduce this background, we
  use the transverse momentum imbalance ($\MPTVEC$) as measured in the
  spectrometer.  This variable is more correlated to the neutrino
  energy and its direction than $\METVEC$ in this class of events. The
  absolute amount of $\MET$ and \mpt, the angle between them,
  the azimuthal angles between $\METVEC$ or $\MPTVEC$ and
  the jet directions, and several other less powerful variables are
  used as inputs to a neural network (NNQCD).  The NNQCD output is
  required to pass a threshold, removing 77\% of the QCD background
  while keeping 91\% of the signal acceptance.

The backgrounds in the MJ analysis due to QCD events and events with
light-flavor jets produced in association with $W$ and $Z$ bosons are
estimated using data in a control region composed of events in which
the $\METVEC$ is aligned with one of the jets.  The observed and
expected event counts for the MJ analysis are given in the \MET+jets
column of Table~\ref{tab:events}.

\begin{table}[ht!]
\caption{Background composition and predicted number of single top events in 3.2~fb$^{-1}$ of
CDF Run II data for the $\ell$+\protect\MET+jets samples (LF, ME, NN, and BDT analyses), 
and 2.1~fb$^{-1}$ of data for the \protect\MET +jets sample (MJ analysis).}
\begin{center}
\begin{tabular}{lrclrcl}
\hline
\hline
Process & \multicolumn{3}{c}{$\ell~+~$\MET~+~jets} &  \multicolumn{3}{c}{$\MET$~+~jets}\\
\hline
$s$-channel signal	              & 77.3   & $\pm$ & 11.2  & 29.6  & $\pm$ & 3.7 \\
$t$-channel signal                    & 113.8  & $\pm$ & 16.9  & 34.5  & $\pm$ & 6.1 \\
$W+HF$         & 1551.0 & $\pm$ & 472.3 & 304.4 & $\pm$ & 115.5 \\
$t\bar{t}$ 	          	      & 686.1  & $\pm$ & 99.4  & 184.5 & $\pm$ & 30.2 \\
$Z$+jets		              & 52.1   & $\pm$ & 8.0   & 128.6 & $\pm$ & 53.7 \\
Diboson    		              & 118.4  & $\pm$ & 12.2  & 42.1  & $\pm$ & 6.7 \\
QCD+mistags                           & 777.9  & $\pm$ & 103.7 & 679.4 & $\pm$ & 27.9 \\
\hline					                       
Total prediction	              & 3376.5 & $\pm$ & 504.9 & 1404 & $\pm$ & 172 \\
\hline		
Observed  		& \multicolumn{3}{c}{3315}  & \multicolumn{3}{c}{1411}	         \\
\hline
\hline
\end{tabular}
\end{center}
\label{tab:events}
\end{table}


After event selection, the samples are dominated by background.  We
further discriminate the signal with multivariate techniques.  Each
multivariate technique defines a function which reduces several
reconstructed quantities for each event into a single output variable
whose distribution can be studied and fit to extract signal and
background contributions.  Validation of the background modeling for
the input variables and output distributions is a crucial step in the
use of multivariate techniques.  We first describe the construction
of our multivariate tools and then the checks we used to prove the
validity of our background model.  The LF, ME, and NN discriminants are
described in~\cite{CDF}.  The BDT discriminant uses a decision
tree method that applies binary cuts iteratively to classify
events~\cite{DT}.  The discrimination is further improved using a
boosting algorithm~\cite{AdaBoost,TMVA}.  The BDT discriminant uses
over 20 input variables. Some of the most sensitive are the
neural-network jet-flavor separator~\cite{Richter:2007zzc}, the
invariant mass of the $\ell\nu b$ system $M_{\ell\nu b}$, the total
scalar sum of transverse energy in the event $H_\mathrm{T}$,
$Q\times\eta$~\cite{cpyuan}, the dijet mass $M_{jj}$, and the
transverse mass of the $W$ boson.

The LFS discriminant uses projective likelihood
functions~\cite{lfcitation} to combine the separation power of several
variables and is optimized to be sensitive to the $s$-channel process.
The subset of the $\ell$+\protect\MET+jets sample with two $b$-tagged
jets is used and consists of 609 events.  The dominant backgrounds are
$W+HF$ and $t{\bar{t}}$ production.  A kinematic fitter is used to
find the most likely resolution of two ambiguities: the $z$-component
of the neutrino momentum and the $b$ jet that most likely came from
the top quark decay.  In addition to the outputs of the kinematic
fitter, other important inputs to the likelihood are the invariant
mass of the two $b$-tagged jets $M_{bb}$, the transverse momentum of
the $b\bar{b}$ system, the leading jet transverse momentum,
$M_{\ell\nu b}$, $H_\mathrm{T}$, and \MET.

The MJ discriminant uses a neural network to combine information from
several input variables.  The most important variables are the
invariant mass of the $\METVEC$ and the second leading jet, the scalar sum
of the jet energies, the $\MET$, and the azimuthal angle between the
$\METVEC$ and the jets.

We combine the LF, ME, NN, BDT, and LFS channels using a
super-discriminant (SD) technique similar to that which was applied
in~\cite{CDF}.  The SD method uses a neural network trained with
neuro-evolution ~\cite{neat} to separate the signal from the
background taking as inputs the discriminant outputs of the five
analyses for each event.  With the super-discriminant analysis we
improve the sensitivity (defined below) by 13\% over the best
individual analysis.  We perform a simultaneous fit over the two
exclusive channels, MJ and SD, to obtain the final combined results.

Before investigating the sample of selected events, we used
background-dominated data control samples to check the modeling of
each input variable as well as the output distributions of each
multivariate discriminant.  For the $\ell~+~$\MET~+~jets analyses the
control samples used are the lepton + $b$-tagged four-jet sample,
which is enriched in $t\bar{t}$ events, and the two- and three-jet
samples in which there is no $b$-tagged jet.  The latter are enriched
in $W$+jets and QCD events with kinematics similar to the $b$-tagged
signal samples and have high statistics, making it possible to observe
that the background model describes the data well over three orders of
magnitude in our output discriminants.  For the MJ analysis, three
control samples are used: in the first sample, the $\METVEC$ is
required to be aligned along one of the jets, and in the second, the
events are required to fail the NNQCD requirement, and in the third, a
lepton is required to be present.  The data distributions in all
control samples are described well by our models for each of the
analysis input variables and a large set of other variables not used
as inputs.  More than two thousand distributions were checked for
evidence of mismodeling.  Small discrepancies were found in the
distributions of the angles between two jets in the untagged lepton +
two-jet sample and the modeling of jets with rapidity greater than
2.4.  These effects are included as systematic uncertainties on the
shape of the background models.

\begin{figure*}[th]
\includegraphics[width=1.0\textwidth]{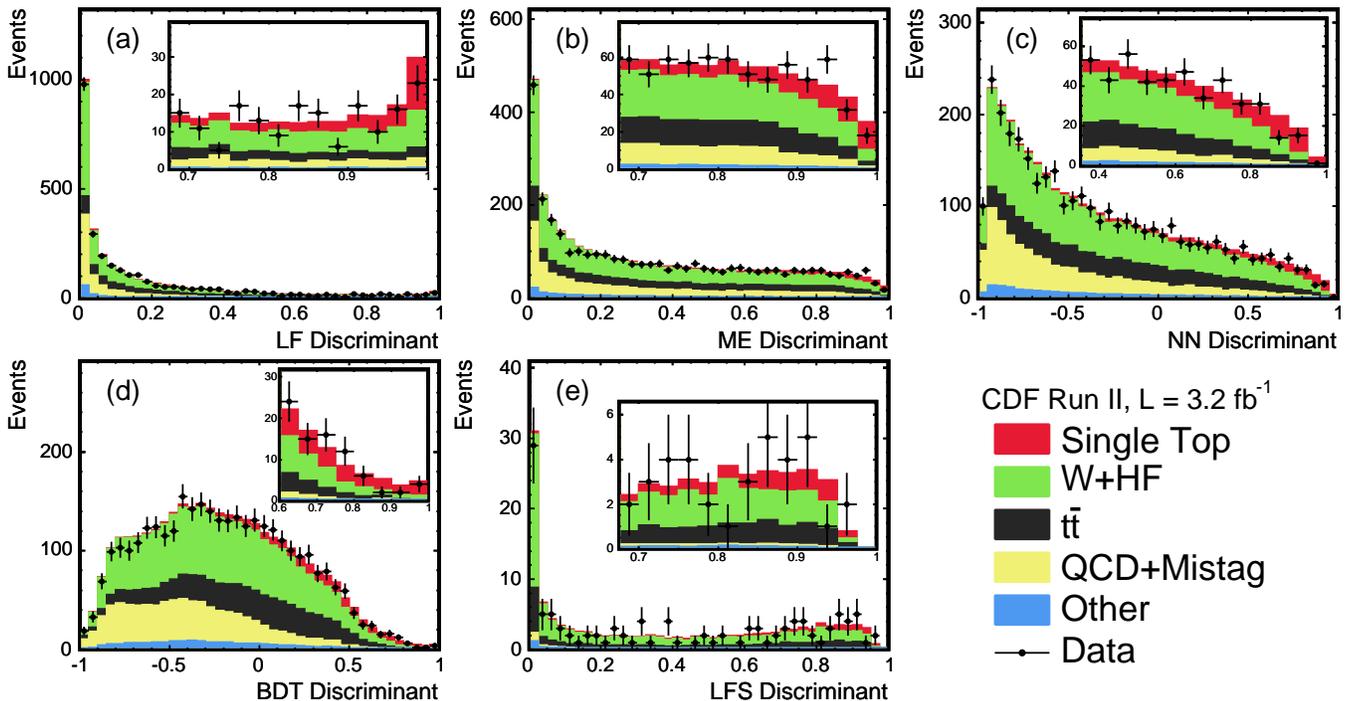}
\caption{Discriminant distributions for the $\ell$+\protect\MET+jets
analyses.  The data are indicated with points, and the predictions are
shown separately for each contribution with stacked histograms.  The
signal expectations shown are the SM predictions.  The insets show the
distributions of the candidate events in the high-discriminant region.}
\label{fig:inputs}
\end{figure*}

\begin{figure*}[th]
\includegraphics[width=1.0\textwidth]{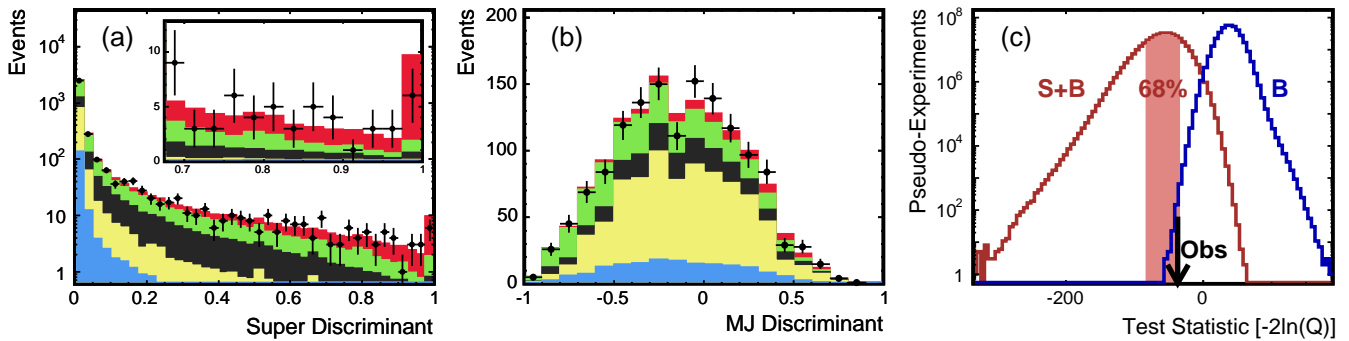}
\caption{Discriminant distributions for the (a) SD, and (b) MJ
analyses (see Fig.~\protect\ref{fig:inputs} for their caption and
legend).  Figure (c) shows the distribution of the likelihood ratio test statistic $-2\ln Q$.
}
\label{fig:superdiscriminant}
\end{figure*}

Figure~\ref{fig:inputs} shows the distributions of the five
$\ell~+~$\MET~+~jets discriminants.  These are combined to give the SD
distribution shown in Fig.~\ref{fig:superdiscriminant} together with
the MJ distribution.  In the rightmost bins, assuming SM production
and decay, the SD has an $s/b$ that exceeds 5.0.  This large $s/b$
significantly reduces our sensitivity to systematic uncertainties
affecting the background. We use the distributions of the SD and MJ
discriminants to extract the measured cross section and the signal
significance.

We measure the single top cross section using a Bayesian binned
likelihood technique~\cite{pdg} assuming a flat prior in the cross
section and integrating the posterior over all sources of systematic
uncertainty.  The background rates are varied within uncertainties,
but are largely constrained by the data in the background-enriched
portions of the SD and MJ discriminant distributions.  Uncertainties
on the shapes of these distributions degrade the extrapolation of
these constraints to more signal-like regions.  The sources of
systematic uncertainties affecting these shapes are discussed below
and are also included in all calculations. The uncertainties assigned
were conservatively chosen to cover the full range of variations
studied.  We quote the measured cross section as the value that
maximizes the posterior likelihood, and use the shortest interval
containing 68\% of the integral of the posterior to set the
uncertainties.  We calculate the significance as a
$p$-value~\cite{pdg}, which is the probability, assuming single top
quark production is absent, that $-2\ln Q=
-2\ln\left(p({\mathrm{data}}|s+b)/p({\mathrm{data}}|b)\right)$ is less
than that observed in the data.  Figure~\ref{fig:superdiscriminant}(c)
shows the distributions of $-2\ln Q$ in pseudoexperiments that assume
SM single top ($S+B$) and also those that assume single top production
is absent ($B$), along with the value observed in data.  The effects
of the systematic uncertainties are included in the pseudoexperiments.
We convert the observed $p$-value into a number of standard deviations
using the integral of one side of a Gaussian function.

All sources of systematic uncertainty are included and correlations
between normalization and discriminant shape changes are considered.
Uncertainties in the jet energy scale, $b$-tagging efficiencies,
lepton identification and trigger efficiencies, the amount of initial
and final state radiation, parton distribution functions,
factorization and renormalization scale, and background modeling have
been explored and incorporated in all individual analyses and the
combination.  We include uncorrelated MC statistical uncertainties in each bin of each discriminant distribution.  A $\pm 2.5$~GeV/$c^2$ uncertainty on the top
quark mass $m_t$ is included in the significance and $|V_{tb}|$ results but quote the dependence on $m_t$ separately in the cross section.

\begin{table}[ht!]
\caption{Results summary for the five correlated $\ell$+\protect\MET+jets 
analyses combined by the SD analysis, 
the SD and the MJ analysis, and the total combination.  The LFS
analysis measures only the $s$-channel production cross section, while the other analyses measure the sum
of the $s$- and $t$-channel cross sections.}
\begin{center}
\begin{tabular}{lccc} \hline\hline
Analysis        & Cross                    & Significance & Sensitivity \\
                & Section (pb)             & (Std. Dev.)         & (Std. Dev.) \\ \hline
LF              & 1.6$^{+0.8}_{-0.7}$      & 2.4          & 4.0    \\
ME              & 2.5$^{+0.7}_{-0.6}$      & 4.3          & 4.9    \\
NN              & 1.8$^{+0.6}_{-0.6}$      & 3.5          & 5.2    \\
BDT             & 2.1$^{+0.7}_{-0.6}$      & 3.5          & 5.2    \\
LFS             & 1.5$^{+0.9}_{-0.8}$      & 2.0          & 1.1    \\ \hline
SD              & 2.1$^{+0.6}_{-0.5}$      & 4.8          & $>5.9$  \\
MJ              & 4.9$^{+2.5}_{-2.2}$      & 2.1          & 1.4    \\ \hline
Combined        & 2.3$^{+0.6}_{-0.5}$      & 5.0          & $>5.9$    \\ \hline\hline
\end{tabular}
\end{center}
\label{tab:results}
\end{table}

Table~\ref{tab:results} lists the measured cross sections and
significances for each of the component analyses and the combination.
The measured cross sections for the five correlated analyses and the
SD are close to each other even though the analyses choose different
input variables and are optimized differently.  We interpret the
excess of signal-like events over the expected background as
observation of single top production with a $p$-value of $3.10\times
10^{-7}$, corresponding to a signal significance of 5.0 standard
deviations. The sensitivity is defined to be the median expected
significance and is in excess of 5.9 standard deviations, assuming the SM
signal cross section.  The most
probable value of the combined $s$-channel and $t$-channel cross
section is $2.3^{+0.6}_{-0.5}$~pb assuming a top quark mass of 175
GeV/$c^2$.  The dependence on the top quark mass is
+0.02~pb/(GeV/$c^2$).  From the cross section measurement at
$m_t=$~175 GeV/$c^2$, we obtain $|V_{tb}|=0.91 \pm 0.11
(\mathrm{stat+syst})\pm 0.07$(theory\cite{harris}) and limit
$|V_{tb}|>0.71$ at the 95\% C.L. assuming a flat prior in $|V_{tb}|^2$
from 0 to 1.  This is the most precise direct measurement of
$|V_{tb}|$ to date.

In summary, we combine six multivariate analysis techniques to  
precisely measure the electroweak single top production cross section 
and the CKM matrix element $|V_{tb}|$.  We have carefully cross-checked our
analysis techniques with data control samples and we assign generous
rate and shape uncertainties to all predictions we use.  Our combined discriminant
allows us to purify a signal sample with $s/b>5.0$ in the most sensitive region,
allowing for a significant outcome in the presence of these 
conservative systematic uncertainties.  We observe single top quark production
for the first time with a significance of 5.0 standard deviations.

\begin{acknowledgments}
We thank the Fermilab staff and the technical staffs of the
participating institutions for their vital contributions. This work
was supported by the U.S. Department of Energy and National Science
Foundation; the Italian Istituto Nazionale di Fisica Nucleare; the
Ministry of Education, Culture, Sports, Science and Technology of
Japan; the Natural Sciences and Engineering Research Council of
Canada; the Humboldt Foundation, the National Science Council of the
Republic of China; the Swiss National Science Foundation; the
A.P. Sloan Foundation; the Bundesministerium f\"ur Bildung und
Forschung, Germany; the Korean Science and Engineering Foundation and
the Korean Research Foundation; the Science and Technology Facilities
Council and the Royal Society, UK; the Institut National de Physique
Nucleaire et Physique des Particules/CNRS; the Russian Foundation for
Basic Research; the Ministerio de Ciencia e Innovaci\'{o}n, and
Programa Consolider-Ingenio 2010, Spain; the Slovak R\&D Agency; and
the Academy of Finland.
\end{acknowledgments}


\begin{thebibliography}{99}
\bibitem{coord} We use a cylindrical coordinate system with its origin in the center 
of the detector, where $\theta$ and $\phi$ are the polar and azimuthal angles, respectively,
and pseudorapidity is $\eta= - \rm ln$ $\rm tan(\theta/2)$. 
The missing $E_T$ ($\METVEC$) is defined by $\METVEC = -\sum_iE_T^i{\hat n}_i, i={\rm calorimeter~tower~number}$,
where ${\hat n}_i$ is a unit vector perpendicular to the beam axis and pointing at the
$i^{th}$ calorimeter tower.  $\METVEC$ is corrected for high-energy muons and also jet energy corrections.
We define $\MET=|\METVEC|$. The transverse momentum $p_T$ is defined to be $p\sin\theta$.
\bibitem{cpyuan} $Q\times\eta$ is
the charge of the lepton times the pseudorapidity of the jet not assigned to be the $b$ from top quark decay; see
  C.~P.~Yuan,
  Phys.\ Rev.\  D {\bf 41}, 42 (1990).
\end{thebibliography}
\end{document}